\documentclass{article}
\newcommand{\mbs}[1]{\mbox{\scriptsize{#1}}}

\renewcommand{\(}{\left (}
\renewcommand{\)}{\right )}
\newcommand{\beq}{\begin{equation}}
\newcommand{\eeq}{\end{equation}}
\newcommand{\beqa}{\begin{eqnarray}}
\newcommand{\eeqa}{\end{eqnarray}}
\newcommand{\scaption}[1]{\caption{\small #1}}

\newcommand{\D}[2]{\Delta_{\mbs{#1}}(\mbs{#2})}
\usepackage{psfig}
\begin{document}
\begin{titlepage}
\begin{flushright}
LU TP 98-5\\
February 28, 1998 \\
\end{flushright}

\vspace{0.8in}
\LARGE
\begin{center}
{\bf Local Routing Algorithms \\
Based on Potts Neural Networks}\\
\vspace{.3in}
\large
Jari H\"akkinen\footnote{jari@thep.lu.se},
Martin Lagerholm\footnote{martin@thep.lu.se},
Carsten Peterson\footnote{carsten@thep.lu.se} and
Bo S\"oderberg\footnote{bs@thep.lu.se}\\
\vspace{0.05in}
Complex Systems Group, Department of Theoretical Physics\\
University of Lund, S\"olvegatan 14A, SE-223 62 Lund, Sweden\\

\vspace{0.4in}

Submitted to {\it IEEE/ACM Transactions on Networking}

\end{center}
\vspace{0.3in}
\normalsize

Abstract:

A feedback neural approach to static communication routing in
asymmetric networks is presented, where a mean field formulation of the
Bellman-Ford method for the single unicast problem is used as a common
platform for developing algorithms for {\it multiple unicast}, {\it
multicast} and {\it multiple multicast} problems.  The appealing
locality and update philosophy of the Bellman-Ford algorithm is
inherited.
For all problem types the objective is to minimize a total connection
cost, defined as the sum of the individual costs of the involved
arcs, subject to capacity constraints. The
methods are evaluated for synthetic problem instances by comparing to
exact solutions for cases where these are accessible, and else with
approximate results from simple heuristics. The computational demand
is modest.

\end{titlepage}

\section{Introduction}
Static routing problems amount to assigning arcs (edges) to a set of
communication requests in a network, such that a total arc cost is
minimized subject to capacity constraints on the arcs.
For a review of such problems and existing routing techniques, see
e.g.~\cite{bert}.
In this paper we develop a family of novel distributed algorithms for
static routing problems with asymmetric links, based on a {\it Potts
neuron} encoding and a mean field ({\bf MF}) relaxation dynamics. Such
approaches have proven powerful in many resource allocation
problems~\cite{gis2}, including cases with a non-trivial
topology~\cite{lager}.

For the relatively simple {\em single unicast} (shortest path)
problem, several fast exact methods exist, scaling polynomially with
the network size, e.g. Bellman-Ford ({\bf BF})~\cite{bell}, Dijkstra
and Floyd-Warshall~\cite{bert}. For other problem types one in
general has to rely on various heuristics~\cite{kou}.
Here, we will exploit a recast of BF in a neural form, with a Potts MF
neuron at each node~\cite{jari1}, as a starting point for approaching
more complex problems, with the appealing local update philosophy 
of BF preserved. For problems with multiple simultaneous requests, 
a separate Potts network is assigned to each request,  
interacting through penalty terms encoding load constraints. 

The {\it multiple unicast} problem, where arcs are to be allocated
simultaneously to several unicast requests, is probably NP-hard,
although we are not aware of any rigorous proof for this. 
%
Preliminary results using the Potts MF approach for this problem 
were reported in~\cite{jari1}.

The generic {\it single multicast} problem, where one message is to 
be sent to several receivers, is known to be NP-complete~\cite{karp}. 
The arc capacities are irrelevant, and the objective is simply to 
minimize the sum of the individual costs of the arcs used.

In a {\it multiple multicast} problem the capacity constraints come
into play, and a load interaction has to be introduced in the
dynamics.

The performance of a Potts MF method for each of these problems is
evaluated by a comparison to different approximate schemes. For
multiple unicast simple heuristics based on BF are used, and for
multicast a Directed Spanning Tree Heuristic ({\bf DSTH}), inspired by
the Minimal Spanning Tree Heuristic~\cite{kou}, is employed.  A
heuristic based on a sequential application of DSTH is used for
multiple multicast. For small enough problems, an exact
Branch-and-Bound ({\bf BB}) method is used to provide an exact
solution.

Despite the global nature of the problems, the implementation of the 
Potts approach is truly local -- when updating the MF neurons for a 
particular node, only information residing at neighbouring nodes is 
needed. Together with a good performance, this represents a key asset 
of the method.

This paper is organized as follows. In Section~\ref{problems} the
problem types are defined and discussed. A section each on the three
problem types then follow, where the corresponding Potts approach is 
described and evaluated. A brief summary can be found in
Section~\ref{SUM}. Algorithmic details on the Potts MF algorithms, the
BB algorithms and a minimal directed spanning tree heuristic are given
in Appendices A, B and C, respectively.

\vfill
\section{Problem Discussion}
\label{problems}

\subsection{Networks}

In the problem types considered, a {\em network} is assumed to be
given, defined by a connected graph of $N$ nodes and $L$
(bidirectional) links, corresponding to $2L$ arcs, each with a
specified cost (arc-length) and capacity.

We will limit our interest to networks with at most one link for every
node pair.  Hence an arc from node $i$ to node $j$ can be
unambiguously labeled by the ordered pair of node labels $(ij)$; this
notation improves readability of formulae and will be used throughout
the paper, with the restriction to linked node pairs being understood.

For an arc $(ij)$, its cost $d_{ij}$ could represent e.g. an actual
cost, or the delay of a signal traveling through the arc, while its
(integer) capacity $C_{ij}>0$ represents the maximum number of
simultaneous signals it can hold.
%

\subsection{Problem Types}

On a given network, a {\em unicast} request is defined by specifying a
{\em sender} node $a$, that is to transmit a message to a {\em
receiver} node $b$.  Similarly, a {\em multicast} request is defined
by specifying a sender node $a$, with a single message aimed at
several receiver nodes $b_i, i=1\ldots B$.

A {\em multiple unicast} problem is then defined by specifying a set
of simultaneous single unicast requests, $a_r,b_r,r=1\dots R$, that
are to be routed, such that the total path length is minimized, without
any arc capacity being exceeded.

In a {\em single multicast} problem, a multicast request is to be
routed through the shortest directed tree that is rooted at the
sender and reaches all the receivers. The length of the tree is
defined as the sum of the arc lengths of the used arcs.

Finally, in a {\em multiple multicast} problem several multicast
requests are given. Each should be routed, such that the sum of the
resulting tree lengths is minimized, while respecting arc capacities.

\subsection{Random Problems}

To gauge the various algorithms for the three problem types, we have
used artificial random problems, defined on a pool of artificially
generated random networks described in Table~\ref{red_table}.

To ensure that a network with $N$ nodes and $L$ links is connected, it
is built by first using $N-1$ links to create a random spanning
tree. Then each remaining link is used to connect a random, previously
unconnected pair of nodes; the two corresponding (oppositely directed)
arcs are independently assigned a random cost in the interval $[0,1]$,
and a random integer capacity in the range $\{1,\dots,6\}$.

When a network of the desired size is chosen from the pool, a random
problem is constructed by generating the desired number of independent
random requests, for the multicast case with the desired number of
receiver nodes.

\subsection{Problem Reduction}

The topology of a network might admit a decomposition into
subnetworks, such that a routing problem reduces to a set of
independent subproblems, each in its own subnetwork.

\begin{figure}[hbt]
\begin{center}
\mbox{\psfig{figure=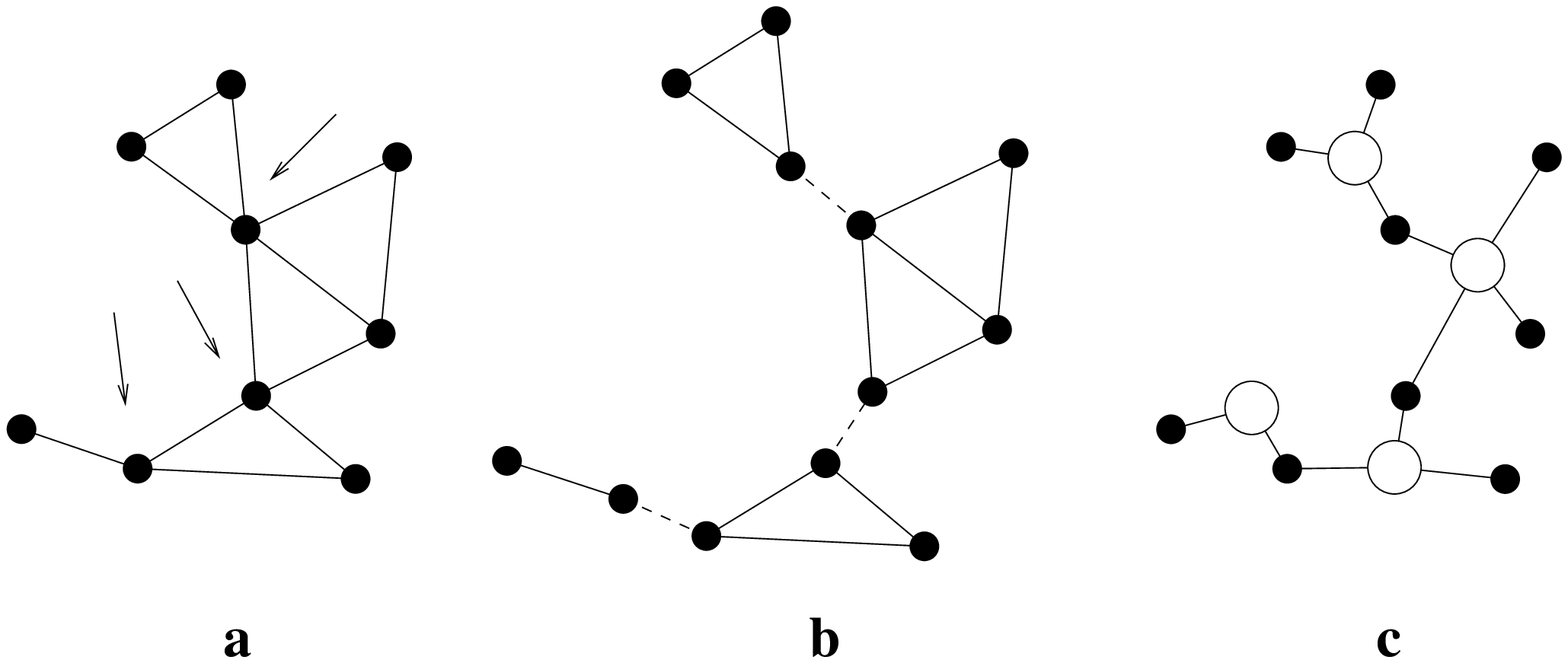,width=10cm}}
\end{center}
\scaption{
 {\bf (a).} A network with the split-nodes marked with arrows.
 {\bf (b).} The decomposition into subnetworks.
 {\bf (c).} The hypertree; every second node (white blobs)
	  represents a subnetwork.}
\label{prepro}
\end{figure}

Every node that is such that its removal will disconnect the network
defines a {\em split-node}. At every such node, the network can be split
in two or more parts, each with its own replica of that node. By splitting
the network at all split-nodes, a tree of subnetworks, connected via the
split-nodes, will be formed; Fig. \ref{prepro} shows an example.

To keep track of the relation between the original problem and the
resulting subproblems, an auxiliary graph, to be referred to as the
{\em hypertree}, can be defined, containing the original nodes as well
as the subnetworks as formal nodes, with links representing the
belonging of a node to a subnetwork, as depicted in
Fig. \ref{prepro}c.  For a routing problem in a reducible network, the
corresponding trivial problem in the hypertree is first solved; its
unique solution determines the decomposition of the given problem into
subproblems.

The complexity of a problem can be taken as the size of the space of
possible routings; this depends on the problem and is in general
difficult to compute.
Instead, noting that the Bellman-Ford algorithm scales as $(N-1) L$,
we will use a rough, but simple and problem-independent measure $Q$ of
the reduction in network complexity, defined as follows.  With $N_s$
and $L_s$ denoting the number of nodes and links in a subnetwork $s$,
we simply define
\beq
	Q
	\equiv \frac {\sum_s \(N_s-1\) L_s} {(N-1) L}
	\equiv \frac {\sum_s \(N_s-1\) L_s} {\sum_s \(N_s-1\) \sum_s L_s}
        \leq 1.
\label{red}
\eeq
Table~\ref{red_table} shows the average reduction factors for the pool
of random networks. The reduction is most important for the exact BB
methods, where the actual time reduction factor can be considerable,
even for $Q\approx 1$.
\begin{table}[tbh]
\begin{center}
\begin{tabular}{|r|r||c|c|}
\hline
$N$ & $L$ &  $<Q>$ \\ 
\hline
  5 &  10  &   1.00  \\ 
 10 &  20  &   0.97  \\ 
 15 &  20  &   0.61  \\ 
 20 &  30  &   0.75  \\ 
 50 & 100  &   0.91  \\ 
 50 & 200  &   1.00  \\ 
100 & 200  &   0.90  \\ 
100 & 400  &   1.00  \\ 
200 & 250  &   0.42  \\ 
\hline
\end{tabular}
\end{center}
\scaption{Sizes chosen for the pool of random networks. 1000 networks
 of each size are generated. Also shown is the average reduction
 factor.}
\label{red_table}
\end{table}
%

\section{Multiple Unicast}
\label{MU}
\begin{figure}[b]
\begin{center}
\mbox{\psfig{figure=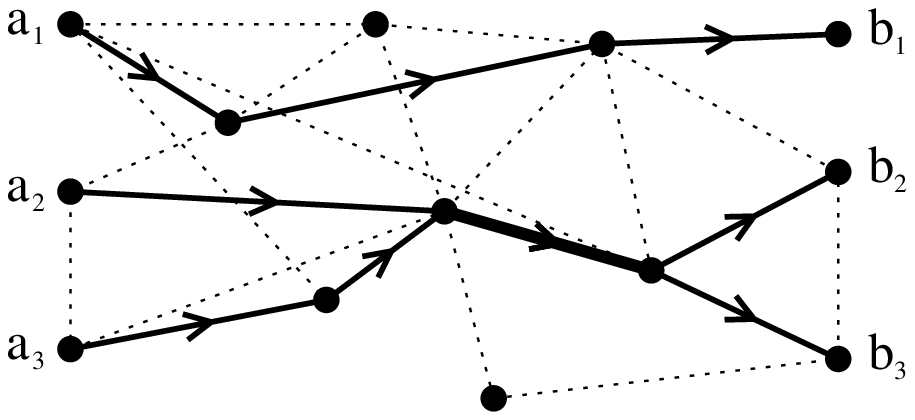,width=6.7cm}}
\end{center}
\scaption{Example of a solution to a 3-request unicast problem.  Dotted
	lines represent unused links, and full lines links that are
	used by the requests.}
\label{pmufig}
\end{figure}

In this section we focus on developing a Potts mean field algorithm
for the multiple unicast problem. This is done in two steps:
In Subsection~\ref{MU_BF} the Bellman-Ford algorithm for the single
unicast problem is recast into a Potts mean field language,
while in Subsection~\ref{MU_PMU} an extension to and tools to handle
the multiple case is described.
%
The solution to a 3-request problem is shown in Fig.~\ref{pmufig}.
%

\subsection{Mean Field Version of the Bellman-Ford Algorithm}
\label{MU_BF}
For a {\it single} unicast, the arc capacities are irrelevant, and the
task is simply to find the shortest path from a sender $a$ to a
receiver $b$. In the BF algorithm~\cite{bell}, this is done by
relaxing, for each node $i$, the estimated shortest path-length $D_i$
to $b$, according to
\beq
\label{BF}
	D_i \to \min_j (d_{ij}+D_j) \equiv \min_j E_{ij} \mbox{, } i \ne b,
\eeq
and keeping track of the chosen neighbours $j$. Note the distinct
philosophy here: Each node $i$ minimizes its own {\em local energy}
$E_{ij}$, rather than all nodes striving to minimize some global
objective function.

Eq.~(\ref{BF}) can be written as
\beq
\label{BF_1}
	D_i = \sum_j v_{ij} E_{ij}
\eeq
in terms of a {\em winner-take-all neuron} ${\bf v}_i$ for every node
$i \neq b$, with components $v_{ij}$ taking the value 1 for the
optimal neighbour $j$, and 0 for the others.

A {\em mean field} (MF) version of BF is obtained by replacing the
discrete winner-take-all neurons by MF Potts neurons, with components
given by
\beq
\label{potts}
	v_{ij} = \frac{e^{-E_{ij}/T}}{\sum_ke^{-E_{ik}/T}} \;,
\eeq
where $T$ is an artificial temperature. The component $v_{ij}$ is to
be interpreted as a {\em probability} for node $i$ to choose neighbour
$j$ as a continuation node, consistently with $\sum_j v_{ij}=1$.

At a non-zero temperature, iteration of~(\ref{BF_1}, \ref{potts}) can
be viewed as a soft (fuzzy) version of the BF algorithm. At high $T$,
the probability is evenly distributed between the available
neighbours, while in the $T \to 0$ limit a winner-take-all dynamics
results,
and the proper BF algorithm is recovered.

\begin{figure}[bt]
\begin{center}
\mbox{\psfig{figure=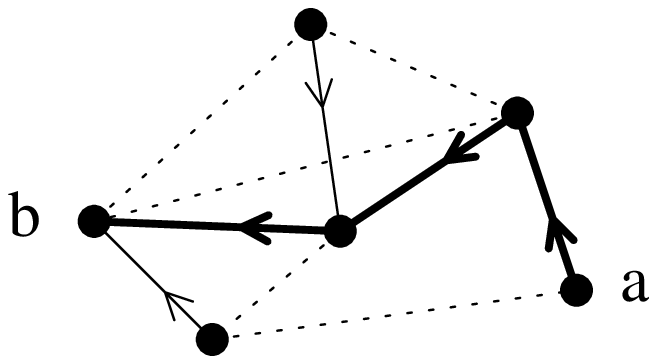,width=5.7cm}}
\end{center}
\scaption{Example of a solution to a single unicast problem. The arcs
 of the emerging tree are shown as solid lines with arrows indicating
 the direction, while unused links are shown dotted. The path from
 {\bf a} to {\bf b} is found by following the arrows; it is marked
 with fat lines.}
\label{BF_tree}
\end{figure}

By starting at a non-vanishing $T$, and slowly letting $T\to 0$
(annealing), the MF neurons will gradually converge to sharp
winner-take-all states, and a spanning tree directed towards $b$ will
emerge (a {\em BF-tree}), consisting of the chosen arcs.
In particular, the cost $D_a$ for the optimal path from sender to
receiver is determined; the path itself is simply extracted from the
spanning tree, c.f. Fig. \ref{BF_tree}.

The resulting MF algorithm will be referred to as {\bf PSU} (Potts
Single Unicast); it is not very interesting in itself, but will serve
as a stepping stone towards MF algorithms for the more difficult
problem types. Note that BF, and thus PSU, will always yield a tree
solution -- loops do not pay.
%

\subsection{The Potts Mean-Field Approach to Multiple Unicast: PMU}
\label{MU_PMU}

The obvious generalization of PSU to a multiple unicast problem is by
having a separate Potts system for each unicast request, labeled $r=1
\dots R$. To enforce the capacity constraints, the local energies
$E^r_{ij}$ will have to be supplemented with a penalty term for arc
overloading. The competition for arcs then destroys the no-loop
guarantee, since the path-choice for one request might block the way
for another. To suppress potential loop-formation, an additional
penalty term is added, and the local energy takes the form
\beq
	E^r_{ij} = d_{ij} + D^r_j
	+ \alpha E^{r,\mbs{load}}_{ij}
	+ \gamma E^{r,\mbs{loop}}_{ij} \;,
\label{E_pmu}
\eeq
where $D^r_j$ is the cost (including load and loop penalties) from
node $j$ to the endnode in request $r$.  Note that the Bellman-Ford
philosophy is kept, only the energy is redefined,
following~\cite{jari1}.

In spite of the loop-suppression term, a neuron might wind up in an
impossible situation, with no good continuation node
available. Following~\cite{jari1} an escape facility is introduced,
enabling an (expensive) emergency route to $b_r$.

To construct the load and loop terms, a {\em propagator} formalism is
employed, following ref.~\cite{lager}; for a more detailed explanation
see~\cite{lager2}. For the Potts system managing the routing of
request $r$, an element of the propagator ${\bf P}^r$ is defined as
\beq
P^r_{ij} = ( ( {\bf 1}-{\bf v}^r )^{-1} )_{ij}  = \delta_{ij} + v^r_{ij}
           + \sum_k v^r_{ik} v^r_{kj} + \sum_{kl}v^r_{ik}v^r_{kl}v^r_{lj}
           + \ldots
\label{P_pmu}
\eeq
It could be interpreted as the (fuzzy) number of paths $i \to j$ in
the BF-tree for request $r$, and becomes integer for $T=0$. It
enables the definition of a probabilistic measure, $F^r_i$, of how
much node $i$ participates in the path $a_r\to b_r$ serving request
$r$,
\beq
	F^r_i \equiv
	\frac{P^r_{a_r i}}{P^r_{ii}} \frac{P^r_{i b_r}}{P^r_{b_r b_r}}
	= \frac{P^r_{a_r i}}{P^r_{ii}} \mbox{ } (\le 1).
\eeq
The simpler form follows from $P^r_{i b_r} \equiv 1$. The desired
penalty terms can now be defined, based on the propagator.

Thus, the load $L^r_{ij}$ on an arc $(ij)$ due to the request $r$ is
given by $L^r_{ij}=F^r_i v^r_{ij}\le 1$. Summing the contributions
from all requests yields the total arc-load, $L_{ij} = \sum_r
L^r_{ij}$.
For a particular request $r$, the overloading of the arc ($ij$) due to
the other requests is given by
\beq
        W(X) \equiv X \Theta(X) \; ,\;\;
        X \equiv L_{ij} - L^r_{ij} - C_{ij} \; ,
\label{W_x}
\eeq
where $\Theta$ is the Heaviside step function. With the arc also
used by $r$, the overloading would increase to $W(X + 1)$, and the
difference will serve as an overloading penalty,
\beq
        E^{r,\mbs{load}}_{ij} = W \left( X + 1 \right) - W \left( X \right) .
\label{Er_load}
\eeq

In addition, the amount of loops introduced by connecting $i$ to $j$
can be expressed as the amount of path from $j$ to $i$, as given by $Y
\equiv P^r_{ji}/P^r_{ii} \;\;(\le 1)$, and we choose as a loop
suppression term
\beq
        E^{r,\mbs{loop}}_{ij} = \frac{Y}{1-Y}.
\label{E_loop}
\eeq

With a separate Potts system for each request, the updating equations
(\ref{BF_1}, \ref{potts}) will be replaced by
\beq
        D^r_i \to \sum_j v^r_{ij} E^r_{ij}
\label{D_upd}
\eeq
with
\beq
        v^r_{ij} = \frac{e^{-E^r_{ij}/T}}{\sum_ke^{-E^r_{ik}/T}}
\label{v_upd}
\eeq
The propagators will be updated in a ``soft'', local manner, relaxing
towards (\ref{P_pmu}):
\beq
        P^r_{im} \to \delta_{im} + \sum_j {v^r_{ij} P^r_{jm}}
        \; , \;\; \mbox{for all } m.
\label{P_upd}
\eeq

The algorithm defined by iterating (\ref{E_pmu}, \ref{D_upd},
\ref{v_upd}, \ref{P_upd}) with annealing in $T$ will be referred to as
{\bf PMU}; contrary to the case for PSU, it does {\em not} correspond
to an exact algorithm in the zero temperature limit.

Note that iterating only at $T=0$ will in general not lead to a good
result, since a choice made for one request could force a sub-optimal
choice for another. By using instead the mean field annealing
technique, the decisions are made in an incremental way, allowing
neurons to gradually form their decisions under the influence of the
emerging decisions of the other neurons.
Note also that the philosophy inherited from the BF algorithm is not
disturbed, all information needed is local to the relevant node $i$
and its neighbours $j$, with each node keeping track of its own row of
${\bf P}^r$.

\subsection{Evaluation of PMU}

The performance of the PMU algorithm is gauged against three other
algorithms.

{\it Independent Bellman-Ford Heuristic -- IBF}\\
Each request is independently solved using BF, disregarding the load
constraints. Thus, the result might be illegal, whereas a legal result
is necessarily the exact minimum. For ``tight'' networks (arc
capacities low in relation the signal density) illegal results are in
general produced.

{\it Sequential Bellman-Ford Heuristic -- SBF}\\
Here, the unicast requests are served in a random order using BF; when
the maximum capacity of an arc is reached, its use is prohibited for
the subsequent requests. This algorithm can be run repeatedly, with
the request order reshuffled in between, until a preset time limit is used
up; then the best result is kept.  It does not always find a legal
result, even to a solvable problem; when it does, it is not
necessarily the minimum.

{\it Branch-and-Bound -- BB}\\
For small enough problems, an exact Branch-and-Bound algorithm,
presented in \ref{a:BB}, is used to find the exact minimum.

The rate of legal results as well as their quality is probed for
network sizes spanning from 5 to 100 nodes, see
Table~\ref{res_pmu_legal}. The computational demand for BB grows very
fast with the network size, therefore a CPU time limit of 5 minutes is
used. For sizes where BB's chances of finding a solution within the
time interval is small, it is not used at all, those entries are marked
with ``---'' in the table. The SBF heuristic is allowed to run for a
slightly longer time than PMU as specified in Table~\ref{res_pmu_legal}.

The quality of the results from an algorithm Y as compared with that of
a reference algorithm X can be measured in terms of the relative excess
path length,
\beq
	\D{X}{Y} = \frac
	{ D_{\mbs{Y}} - D_{\mbs{X}} }
	{ D_{\mbs{X}} },
\label{delta_XY}
\eeq
where $D_{\mbs{X}}$ is the total path costs resulting from X. Mean
values for problems where both algorithms gave legal results are
presented in Table~\ref{res_pmu_legal}.

From Table~\ref{res_pmu_legal} one finds as could be expected that not
all problems were solvable, and that PMU and SBF gave rise to
approximately the same number of legal results. Problem instances with
a legal IBF result can be regarded as easy: The minimum coincides with
that of the unconstrained problem. From Table \ref{res_pmu_legal}, one
finds that in a region of tight problems PMU is doing slightly better
than SBF, whereas the oppposite is true in the other end.
\begin{table}[bth]
\begin{center}
\begin{tabular}{|r|r|r||r|r|r|r|r|c|c|c|c|c|r|}
\cline{4-7} \cline{9-13} \multicolumn{3}{c|}{} &
\multicolumn{4}{c|}{legal results} & \multicolumn{1}{c|}{} &
\multicolumn{2}{c|}{$<$CPU$>$} & \multicolumn{3}{c|}{$<\D{X}{PMU}>$} \\
\hline $N$ & $L$ & $R$ & PMU & BB & IBF & SBF &
$\mbox{T}_{\mbs{BB}}$ & PMU & SBF   & BB & IBF & SBF\\ \hline
  5 &  10  &   5 & 1000 & 1000 & 713 & 1000  &  0 & 0.1  & 1.0        & 0.002 & 0.000 &  0.00049  \\%
 10 &  20  &   5 &  994 & 1000 & 705 &  994  &  0 & 0.2  & 1.0        & 0.003 & 0.000 &  0.00247  \\%
 10 &  20  &  10 &  973 &  954 & 289 &  973  & 19 & 0.4  & 1.0        & 0.005 & 0.000 &  0.00341  \\%
 10 &  20  &  15 &  936 &   93 &  91 &  933  &839 & 0.7  & 1.0        & 0.004 & 0.001 &  0.00519  \\%
 15 &  20  &  15 &  460 &  --- &  39 &  484  & -- & 0.6  & 5.0        &  ---  & 0.000 &  0.00654  \\%
 20 &  30  &  20 &  508 &  --- &   5 &  496  & -- & 1.6  & 5.0        &  ---  & 0.000 & -0.00364  \\%
 50 & 200  &  50 &  996 &  --- &   0 &  997  & -- &  53  & 100        &  ---  &  ---  & -0.00006  \\%
100 & 200  & 100 &  297 &  --- &   0 &  313  & -- & 268  & 300        &  ---  &  ---  & -0.03903  \\%
100 & 400  & 100 &  988 &  --- &   0 &  990  & -- & 515  & 600        &  ---  &  ---  & -0.02229  \\%
\hline
\end{tabular}
\end{center}
\scaption{{\bf Multiple unicast}. Number of legal results and CPU time
 used fro the different algorithms, and a quality
 comparison. $\mbox{T}_{\mbs {BB}}$ is the number of instances for
 which BB was prematurely terminated. 1000 instances of each problem
 size are probed. CPU is given in seconds on an DEC Alpha 250. The
 quality of the PMU results is gauged by comparing to BB, IBF and SBF
 respectively, for instances where both algorithms produced legal
 result (for BB when it found a solution within the time limit); a
 negative $\D{X}{PMU}$ indicates that PMU is superior to X.}
\label{res_pmu_legal}
\end{table}
%

\section{Single Multicast}
\label{SM}

An attempt to solve a multicast problem should yield a multicast tree
({\bf MT}), defined as a directed tree spanning the set of sources $S$
(the sender and the receivers), rooted at the sender. A minimal MT is
of course desired; arc capacities are irrelevant.

\subsection{The Potts Mean-Field Approach to Single Multicast: PSM}
\label{SM_PSM}

The update philosophy of PSU for a single unicast problem is based on
each node minimizing its estimated distance to the end node. For the
multicast problem we instead adopt the strategy that each node $i$
attempts to minimize the distance to its {\em forward MT}, defined as
the partial MT that is spanned by those sources connecting to the
sender via paths {\em not} passing node $i$. Apart from this
modification, the development of PSM closely follows PSU.

\begin{figure}[bt]
\begin{center}
\mbox{\psfig{figure=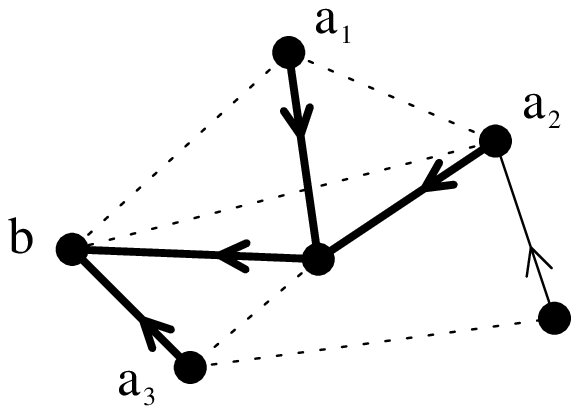,width=5.0cm}}
\end{center}
\scaption{Example of an MT solving a single multicast problem in the
 transposed network. The nodes $a_i$, $i=1,2,3$, are receivers and $b$
 the sender. The arcs of the emerging BF-tree are shown as solid lines
 with arrows, with the subset defining the MT marked with fat
 lines. Unused links are shown dotted.}
\label{MT_tree}
\end{figure}

PSU gives a tree directed {\em towards} the root, a BF-tree. We
want the opposite direction, equivalent to transposing $d_{ij}$ and
$C_{ij}$ (since the network is asymmetric).
We will therefore work entirely in the transposed network, where a
legal result corresponds to an MT directed {\em towards} $b$, now
corresponding to the sender-node. In PSU a single path was extracted
from the BF-tree; in the multicast case, a subtree corresponding to an
MT should be extracted.
For an example of a solution to a multicast problem in the transposed
network, see Fig.~\ref{MT_tree}.

Let $D^{ia}$ denote the estimated distance from node $i$ to the
(fuzzy) path originating from a source node $a \in S$. It is handled
by node $i$ and calculated via the neighbours (propagated) according
to
\beq
\label{Dai}
	D^{ia} = F_{ai} \sum_j v_{ij} \( d_{ij} + D^{ja} \)
	\mbox{, where }
	F_{ai} = 1-\frac{P_{ai}}{P_{ii}} \;.
\eeq
The factor $F_{ai}\le 1$ is a measure to what extent the path from $a$ to
$b$ avoids node $i$. This factor drives $D^{ia}$ to measure the
distance from $i$ to the path from $a$.

For sharp paths, the distance $\tilde{D}_{ij}$ to the forward MT of
node $i$ via a neighbour node $j$, could then be expressed as
$d_{ij}+\min_{a\in S} D^{ja}$, restricted to $a$ with $F_{ai} = 1$
($i$ not in its path). For fuzzy paths, one could use $d_{ij}+\min_{a
\in S} D^{ja}/F_{ai}$, where a source $a$ with $i$ in its path is
penalized with a large factor. Instead of entirely using $D^{ja}$ for
the best source $a$, however, we will use a weighted value in the MF
spirit, given by
\beq
	\tilde{D}_{ij} = d_{ij} + \frac
	{ \sum_{a\in S} D^{ja}	e^{-D^{ja}/\(F_{ai}\kappa T\)} }
	{ \sum_{a\in S}		e^{-D^{ja}/\(F_{ai}\kappa T\)} } \; ,
\label{u_tD_upd}
\eeq
where $\kappa$ is a suitably tuned temperature ratio.

The neurons $v_{ij}$, managing the choice of neighbour, are
updated according to~(\ref{potts}), with the local energy redefined as
\beq
	E_{ij} = \tilde{D}_{ij} + \gamma E^{\mbs{loop}}_{ij}.
\label{E_psm}
\eeq
where $E^{\mbs{loop}}_{ij}$ is defined as in~(\ref{E_loop}) but for
one request, i.e. neglecting the index $r$\footnote{Using the loop
term will penalize loops immediately; this improves the final
result.}.
A load term is not needed; arcs not capable of hosting at least one
signal are ignored. Likewise, the propagator is updated in the same
manner as for PMU, i.e.~(\ref{P_upd}) neglecting the index $r$. The
resulting algorithm will be referred to as {\bf PSM}.

\subsection{Evaluation of PSM}

In order to evaluate the performance of PSM
we have developed a {\it Directed Spanning Tree Heuristic, DSTH,} as a
heuristic adaptation to asymmetric networks of the Minimal Spanning
Tree Heuristic, MSTH~\cite{kou}. Algorithmic details are given in
\ref{a:DSTH}. For small enough problems, a BB algorithm, described in
\ref{a:BB}, is used to find a minimal MT.

Table~\ref{res_smulticast_legal} shows the number of legal results and
the average CPU consumption for the different algorithms, and a
performance comparison according to (\ref{delta_XY} based on the
MT-cost.  A bar (``---'') marks unexplored entries.

From Table~\ref{res_smulticast_legal} one finds that to all except 2
of the 9000 probed problems the PSM algorithm gave legal results, with
an average quality consistently superiour to that of DSTH.
\begin{table}[hbt]
\begin{center}
\begin{tabular}{|r|r|r||r|r|r|c|c|c|c|c|}
\cline{4-11}
\multicolumn{3}{c|}{}  & \multicolumn{3}{c|}{legal results} & \multicolumn{3}{c|}{CPU time} & \multicolumn{2}{c|}{$<\D{X}{PSM}>$}\\
\hline
$N$ & $L$ & $B$ & PSM & BB & DSTH & PSM & BB & DSTH & BB & DSTH \\
\hline
  5 &  10  & 4   & 1000 & 1000 & 1000 & 0.07  & 0.00 & 0.00 & 0.004  & -0.062 \\
 10 &  20  & 4   & 1000 & 1000 & 1000 & 0.19  & 0.00 & 0.00 & 0.009  & -0.056 \\
 10 &  20  & 9   &  999 & 1000 & 1000 & 0.27  & 0.01 & 0.00 & 0.011  & -0.069 \\
 15 &  20  & 14  & 1000 & 1000 & 1000 & 0.24  & 0.01 & 0.00 & 0.012  & -0.047 \\
 20 &  30  & 15  & 1000 & 1000 & 1000 & 0.59  & 0.07 & 0.00 & 0.017  & -0.056 \\
 50 & 200  & 40  & 1000 &  --- & 1000 & 12.8  &  --- & 0.03 &  ---   & -0.117 \\
100 & 200  & 99  & 1000 &  --- & 1000 & 24.8  &  --- & 0.19 &  ---   & -0.097 \\
100 & 400  & 99  & 1000 &  --- & 1000 & 69.9  &  --- & 0.29 &  ---   & -0.127 \\
200 & 250  & 150 &  999 &  --- & 1000 & 30.3  &  --- & 0.37 &  ---   & -0.050 \\
\hline
\end{tabular}
\end{center}
\scaption{{\bf Single multicast}. The number of legal results, average
 consumed CPU time and the relative quality of the results from the
 different algorithms. 1000 instances of each problem size are
 probed. Same notation as in Table~\ref{res_pmu_legal}; $B$ denotes
 the number of receivers in each multicast.}
\label{res_smulticast_legal}
\end{table}
%

\section{Multiple Multicast}
\label{MM}

A Potts algorithm for the multiple multicast problem is constructed in
an obvious way, by extending PSM in analogy to the extension of PSU to
PMU (Subsection~\ref{MU_PMU}).  Evaluation of the resulting algorithm
is done by comparing to BB and a sequential DSTH heuristic.

\subsection{The Potts Mean-Field Approach to Multiple Multicast: PMM}
\label{MM_PMU}

As in PMU, a separate Potts system ${\bf v}^r$, with a corresponding
propagator ${\bf P}^r$, is introduced for each of the $R$ multicast
requests $r$. The load constraints are again relevant; the load on an
arc from the (fuzzy) MT of a request $r$ is calculated as a ``fuzzy
OR'' over the corresponding source paths,
\beq
	L^r_{ij} = 1 - \prod_{a \in S^r}
	(1 - \frac{P^r_{a i}}{P^r_{ii}} v^r_{ij} ) \; \le 1 \; .
\eeq
where $S^r$ denotes the set of nodes defined by the sender and the
receivers in request $r$.
A penalty term is formed according to~(\ref{W_x}, \ref{Er_load}) with
the total arc load given by $L_{ij} = \sum_r L^r_{ij}$. For a loop
penalty term, (\ref{E_loop}) is used without modification.
The escape facility is defined and used in the same way as in PMU. In
analogy with Subsection~\ref{SM_PSM}, but now for each request $r$,
each node $i$ attempts to minimize the distance to the corresponding
forward MT with respect to the choice of neighbour $j$. For a fixed
$j$, this distance is estimated as
%
\beq
   \tilde{D}^r_{ij} = d_{ij} + \sum_{a \in S_r} D^{rja}
   \frac{
	e^{-D^{rja} / \(F^r_{ai}\kappa T\)}
   }{
	\sum_{b \in S_r}
	e^{-D^{rjb} / \(F^r_{bi}\kappa T\)}
   }\; ,
\label{tD_and_u_upd}
\eeq
where $D^{ria}$ is the distance from $i$, along the fuzzy BF-tree
serving request $r$, to the path from one of its source nodes $a$; it
is updated as
\beq D^{ria} = F^r_{a_ri} \sum_j v^r_{ij} \( d_{ij} + D^{rja} \) \; .
\label{PMM_D_upd}
\eeq
The local energy corresponding to request $r$ thus takes the form
\beq
	E^r_{ij} = \tilde{D}^r_{ij}
	+ \alpha E^{r,\mbs{load}}_{ij}
	+ \gamma E^{r,\mbs{loop}}_{ij} \; .
\label{E_pmm}
\eeq
The neurons and the propagators are updated as in
Section~\ref{MU_PMU}, i.e. using the equations~(\ref{v_upd})
and~(\ref{P_upd}), respectively. The resulting algorithm will be
referred to as {\bf PMM}.

\subsection{Evaluation of PMM}

The performance of the PMM approach is gauged using a Branch-and-Bound
algorithm, presented in \ref{a:BB} and a {\it Sequential Directed
Spanning Tree Heuristic, SDSTH}, an extension of the DSTH algorithm
described in~\ref{a:DSTH}: A DSTH is applied to each multicast request
in turn, blocking arcs having reached their capacity.  The SDSTH is
used in the same way as SBF was used in the multiple unicast case; it
is allowed to run for the time specified in
Table~\ref{res_mmulticast_legal}.

The results for the different algorithm are shown in
Table~\ref{res_mmulticast_legal}.
The PMM heuristic finds approximately the same number of legal results
as SDSTH ($\approx $ 1\% less), with a consistently better quality.
\begin{table}[hbt]
\begin{center}
\begin{tabular}{|r|r|r|r||r|r|r|r|c|c|c|c|}
\cline{5-7}
\cline{9-12}
\multicolumn{4}{c|}{}  & \multicolumn{3}{c|}{legal results} & \multicolumn{1}{c|}{} & \multicolumn{2}{c|}{CPU usage}
& \multicolumn{2}{c|}{$<\D{X}{PMM}>$}\\
\hline
$N$ & $L$ & $R$ & $B$ & BB  & SDSTH & PMM & $\mbox{T}_{\mbs{BB}}$ & PMM & SDSTH & BB & SDSTH \\
\hline
  5 &  10  &   5 &   4 &  989 &    998 &  998 &    9 & 0.43 & 1    &  0.008 &    -0.064  \\
 10 &  20  &  10 &   4 &    0 &    755 &  700 &  722 & 2.03 & 5    &   ---  &    -0.005  \\
 15 &  20  &   4 &   5 &  --- &    565 &  596 &  --- & 0.74 & 5    &   ---  &    -0.017  \\%
 20 &  30  &   5 &   5 &  --- &    673 &  671 &  --- &  2.1 & 5    &   ---  &    -0.020  \\%
 20 &  30  &   5 &  10 &  --- &    267 &  270 &  --- &  5.3 & 6    &   ---  &    -0.019  \\
 50 & 100  &   5 &  25 &  --- &    321 &  296 &  --- & 27.5 & 40   &   ---  &    -0.053  \\
 50 & 200  &  10 &  25 &  --- &    881 &  871 &  --- &  100 & 150  &   ---  &    -0.118  \\
\hline
\end{tabular}
\end{center}
\scaption{{\bf Multiple multicast}. The number of legal results and
 average consumed CPU time for the different algorithms, and a quality
 comparison. 1000 instances of each problem size are probed. Same
 notation as in
 Tables~\ref{res_pmu_legal}~and~\ref{res_smulticast_legal}.}
\label{res_mmulticast_legal}
\end{table}

\section{Summary}
\label{SUM}

A family of Potts mean field feedback artificial neural network
algorithms is developed and explored for artificial multiple unicast
and single as well as multiple multicast routing problems.

In order to handle loads and loops for ``fuzzy'' paths, and other
probabilistic measures, a propagator formalism is used.

The Potts approach is local for all the problem types, with all
information needed by a node for an update residing at this
node and its neighbours. This attractive feature, inherited from the
single unicast Bellman-Ford algorithm, facilitates a distributed
implementation.

The Potts algorithms are compared to competitive heuristics, and gauged
against exact methods whenever feasible, with encouraging results.
For large problems they perform consistently better than the 
other heuristics.

The CPU consumption is proportional to the product of the number of 
requests, receivers, nodes and links.

Another neural network method~\cite{boyan} has been proposed for the
multiple unicast problem; in contrast to our approach it is aimed at
dynamical problems. However, in the static limit it reduces to the
independent BF approach, which is used for comparisons in this work.

\appendix
\renewcommand{\thesection}{Appendix \Alph{section}}
\renewcommand{\theequation}{\mbox{A}\arabic{equation}}
\setcounter{equation}{0}

\newcommand{\AlgBox}[2]{
\begin{minipage}[t]{15cm}
{\large\bf #1}\\[3mm]
\framebox[15cm][l]{
\begin{minipage}{14.5cm}#2
\end{minipage}}\end{minipage}
}
\vfill

\section{PMM -- Algorithmic Details}
\label{a:Pottsalg}
The temperature $T$ is assigned a tentative initial value of $T_0 =
150$. Until $(\Delta v)^2 < 0.1$ after one iteration, with
\beq
\label{B1}
	(\Delta v)^2 = \frac{1}{R (N-1)} \sum_{rij}(\Delta v^r_{ij})^2
	= \frac{1}{{R (N-1)}} \sum_{rij}(v^r_{ij}(t+1)-v^r_{ij}(t))^2 > 0.1
\eeq
the system is reinitialized with $T_0 \rightarrow 2 T_0$.

Each Potts neuron ${\bf v}^r_i$ is initialized by assigning equal
values, with 1\% random variation,
to its components $v^r_{ij}$,
consistent with a unit component sum. Subsequently, $P^r_{ij}$,
$L_{ij}$ and $L^r_{ij}$ are initialized consistently with the neuron
values. $D^{ria}$ is initialized in an approximately consistent way
for each source node $a$, with $D^{rba}=0$.

The following iteration is repeated, until one of the termination
criteria (see below) is fulfilled:
\\[3mm]
\AlgBox{Iteration}
{
  \begin{itemize}
    \item For each single request, $r$, do:
    \begin{enumerate}
      \item Compute the load $L^r_{ij}$ and subtract from $L_{ij}$ for all arcs ($ij$).
      \item For each node $i\ne b$, do:
      \begin{enumerate}
	\item Calculate $E^r_{ij}$ for each neighbour $j$, using (\ref{E_pmm}).
	\item Update $v^r_{ij}$ for each neighbour $j$, using (\ref{v_upd}).
	\item Update $D^{ria}$ for each source $a$, using (\ref{PMM_D_upd}).
	\item Update $P^r_{ij}$ for all nodes $j$, using (\ref{P_upd}).
      \end{enumerate}
     \item Compute the new load $L^r_{ij}$ and add to $L_{ij}$ for all arcs ($ij$).
    \end{enumerate}
    \item Decrease the temperature: $T \to kT$.
  \end{itemize}
}
\\[3mm]
The updating process is terminated when the neurons have almost
converged to sharp 0/1 states, or if it is obvious that they will not
(signaled by a very low $T$), as defined by the criterion
\beq
	\frac{1}{R (N-1)} \sum_{rij} (v^r_{ij})^2 > 1-10^{-6} \;\;\mbox{AND}\;\;
	\max_{rij}(\Delta v^r_{ij})^2 < 10^{-9} \;\; \mbox{OR} \;\; T = 0.00001 .
\eeq

We have consistently used $k = 0.95$ as the annealing rate. The
coefficients $\alpha$ and $\gamma$ in~(\ref{E_pmu}, \ref{E_psm},
\ref{E_pmm})~are chosen as 5 and 0.1 respectively. $\kappa$
in~(\ref{u_tD_upd},~\ref{tD_and_u_upd}) are chosen as 10.

For the {\em single} multicast case, the $r$-loop disappears;
furthermore, the load constraint is irrelevant, so the calculation of
loads is unnecessary and the energy should be calulated using the
simpler formula in~(\ref{E_psm}). For the multiple {\em unicast} case,
the energy should be calculated according to~(\ref{E_pmu}) and ${\bf
D}^r$ as~(\ref{D_upd}).
%

\section{Branch-and-Bound Algorithms}
\label{a:BB}

As exact solvers for small enough problems, two BB algorithms have
been developed, one for single multicasts, the other for multiple
problems.

The {\bf single multicast BB} finds a minimal MT, by starting from a
tree embryo containing only the sender, and recursively adding paths
to each receiver. Each path is recursively constructed, one arc at a
time, starting from the receiver end; it is forced to avoid itself,
and is complete when the existing partial tree is reached. The lowest
MT cost so far is used in every step as a bound to avoid unnecessary
searching.\footnote{Optionally, an initial cost bound (obtained e.g. from a
heuristic) can be specified to further narrow down the search.}

The {\bf multiple multicast BB} is based on an initial recursive
generation of the entire set of possible MT's for each request
separately. This is done as in the single multicast BB, but no bound
is used.
For each MT, its cost as well as data on its arc usage are
stored. Each MT list is then sorted in increasing cost order.
\\ The proper BB part then consists in a recursive traversal of the space
of combinations of one MT for each request, taken from its respective
list. Large sets of combinations are avoided before completion, based
either on exceeding the lowest total cost found so far (particularly
effective due to the sorted lists), or on arc overloading.

The {\bf multiple unicast BB} is contained in the above as an obvious
special case, where an MT is a simple path from the sender to the
receiver.

\section{The Directed Spanning Tree Heuristic}
\label{a:DSTH}
The single multicast heuristic DSTH consists in the following steps:
\begin{enumerate}
\item
Find the shortest paths between all pairs of nodes in $S$ (the set of
sender and receivers), yielding a distance matrix $D$ restricted to
$S\times S$.
\item
Interpret the elements of $D$ as arclengths in an auxiliary complete
graph $G'_s$ spanning $S$.
\item
Find an approximately minimal directed tree $T'_s$, spanning $G'_s$
and rooted at the sender node, by starting with a tree containing the
sender node only, and repeatedly connecting the node with the shortest
arc to the existing tree.
\item
Obtain a subnetwork $G_s$ of the original network $G$ by expanding the
arcs of $T'_s$ in terms of the corresponding shortest paths in $G$.
\item
Find (as in point 3) an approximately minimal directed tree $T \subseteq
G_s$, rooted at the endnode.
\item
Finally, obtain a proper MT by removing from $T$ branches not needed
to span $S$.
\end{enumerate}



\end{document}